\documentclass[nohyperref]{article}

\usepackage{microtype}
\usepackage{graphicx}
\usepackage{subfigure}
\usepackage{booktabs}
\usepackage{siunitx}
\usepackage{etoolbox}
\robustify\bfseries
\usepackage{multirow}
\usepackage{hyperref}

\usepackage[accepted]{icml2022}

\usepackage{amsmath}
\usepackage{amssymb}
\usepackage{mathtools}
\usepackage{amsthm}

\usepackage[capitalize,noabbrev]{cleveref}

\icmltitlerunning{Uncertainty Quantification for Atlas-Level Cell Type Transfer}

\begin{document}

\twocolumn[
\icmltitle{Uncertainty Quantification for Atlas-Level Cell Type Transfer}

\icmlsetsymbol{equal}{*}

\begin{icmlauthorlist}
\icmlauthor{Jan Engelmann}{helmholtz,equal}
\icmlauthor{Leon Hetzel}{helmholtz,tum,equal}
\icmlauthor{Giovanni Palla}{helmholtz,lifescience,equal}
\icmlauthor{Lisa Sikkema}{helmholtz,lifescience}
\icmlauthor{Malte Luecken}{helmholtz}
\icmlauthor{Fabian Theis}{helmholtz,tum,lifescience}
\end{icmlauthorlist}

\icmlaffiliation{helmholtz}{Institute of Computational Biology, Helmholtz Center Munich, Germany}
\icmlaffiliation{lifescience}{TUM School of Life Sciences Weihenstephan, Technical University of Munich, Germany}
\icmlaffiliation{tum}{Department of Mathematics, TU Munich, Germany}

\icmlcorrespondingauthor{Fabian Theis}{fabian.theis@helmholtz-muenchen.de}

\icmlkeywords{Machine Learning, ICML, single-cell genomics, cell type classification}

\vskip 0.3in
]

\begin{NoHyper}\printAffiliationsAndNotice{\icmlEqualContribution}\end{NoHyper}

\begin{abstract}
Single-cell reference atlases are large-scale, cell-level maps that capture cellular heterogeneity within an organ using single cell genomics. Given their size and cellular diversity, these atlases serve as high-quality training data for the transfer of cell type labels to new datasets. Such label transfer, however, must be robust to domain shifts in gene expression due to measurement technique, lab specifics and more general batch effects. This requires methods that provide uncertainty estimates on the cell type predictions to ensure correct interpretation. Here, for the first time, we introduce uncertainty quantification methods for cell type classification on single-cell reference atlases. We benchmark four model classes and show that currently used models lack calibration, robustness, and actionable uncertainty scores. Furthermore, we demonstrate how models that quantify uncertainty are better suited to detect unseen cell types in the setting of atlas-level cell type transfer.
\end{abstract}

\section{Introduction}
\label{introduction}
Single-cell genomics enables the characterization of cellular states at unprecedented scale and resolution. This has led to the generation of diverse single-cell RNA sequencing (scRNA-seq) datasets that capture cellular heterogeneity at the tissue and organ level. To combine multiple studies into a single representation of a human organ, consortia such as the Human Cell Atlas \cite{Regev2017-em} are leading the generation of \emph{integrated scRNA-seq atlases}. Such atlases are built by integrating multiple datasets across samples, individuals, and technologies, and are deemed to comprehensively capture cellular variation in the given tissue or organ. Recently, the first such comprehensive representation of the human lung was introduced in the Human Lung Cell Atlas (HLCA) \cite{Sikkema2022.03.10.483747}.

Next to their potential for discovery, integrated atlases can rapidly accelerate the analysis of new datasets in a harmonised way. Most notably, this is achieved by transferring cell type labels from the reference atlas to new, unannotated data from the same tissue. This task relies on cell type classifiers which are trained on the latent representation and expert cell labels of the integrated atlas. In order to analyze a new dataset, one ``projects"  the query data to the reference's latent space and applies the  classifiers to infer cell type labels. The encoding of the query data is based on heavily regularized batch effect correction approaches to match the reference and is therefore not guaranteed to align perfectly \cite{Lotfollahi2021-mb}. Note that the resulting misalignment naturally becomes greater when the projected data is comprised of samples from a different biological condition, e.g., disease or development, or contains unseen cell states. 

The scenarios described above result in variable degrees of dataset ``shifts", complicating the transfer of cell type labels. In addition, it is paramount to distinguish shifted latent representations due to unresolved technical batch effects from shifts due to biological differences between the datasets, e.g., new cell types or disease-related cell states. Such a distinction can be accomplished with a classifier that provides a measure of uncertainty, which yields actionable insights for downstream tasks.

In machine learning research data shifts, and how classifiers can overcome them, have been widely studied in \emph{uncertainty quantification} \cite{Gal2016UncertaintyID,Gawlikowski-unc}. Crucially, model predictions should be reliable, i.e. calibrated, as they are used by data analysts to draw key interpretations from the data which influence further downstream analyses. Cell type classifiers should maintain these properties across dataset shifts while remaining sensitive to subtle biologically driven shifts in gene expression. Finally, they should be transparent to the analyst, by providing an accurate measure of confidence on the prediction task.

In this work, we evaluate uncertainty quantification methods for atlas-level cell type transfer. We find that models that quantify uncertainty are more robust, better calibrated, and provide high quality uncertainty measures that enable them to identify unseen cell types in the projected data.

\section{Modelling and evaluation of uncertainty}
\label{methods}
\subsection{Dataset and model choice}
The HLCA dataset, on which all evaluations are performed, consists of a core of 14 datasets including 166 scRNA-seq healthy tissue samples of the human respiratory system from 107 individuals \cite{Sikkema2022.03.10.483747}, see Figure\,\ref{supfig:hlca_umap} in appendix\,\ref{sec:appendix} for UMAPs. Batch effects between datasets were removed using scANVI with dataset as the batch covariate \cite{Xu2021-wi}. All cells in the integrated atlas were re-annotated using iterative clustering, data-driven marker gene detection, and a consensus manual annotation from 6 lung experts, leading to 58 consensus cell type labels \cite{Sikkema2022.03.10.483747}. The resulting training data consists of $\sim$580\,000 cells where each cell is represented by a 30 dimensional embedding and has a unique cell type.

As we focus on the models' ability to provide uncertainty estimates, we consider two baseline models as well as two state-of-the-art classifiers:

\begin{itemize}
    \item \textit{Weighted K-nearest neighbor} (WKNN) classifier which is used in the scArches method \cite{Lotfollahi2021-mb} and in the HLCA \cite{Sikkema2022.03.10.483747}.
    \item \textit{Random Forest} (RF) classifier from scikit-learn \cite{DBLP:journals/corr/abs-1201-0490}.
    \item \textit{Deep kernel learning with spectral normalized residual network classifier} (DKL) \cite{Amersfoort, wilson16}
    \item \textit{Multi-input multi-output classifier} (MIMO)  \cite{havasi2021training}, with 3 or 8 subnetworks.
\end{itemize}

Note that we evaluate the models' uncertainty only on the latent embedding and do not have access to the network used for data integration. This makes methods like scANVI \cite{Xu2021-wi} for uncertainty prediction inapplicable.

Uncertainty estimates can be divided into \emph{predictive}, \emph{model}, and \emph{data uncertainty} \cite{Gal2016UncertaintyID,Gawlikowski-unc}. Data uncertainty entails label and measurement noise that is inherent to the data, while model uncertainty measures the uncertainty of the model's predictions due to parameter uncertainty. An application of model uncertainty is OOD detection, because it is higher for OOD cells than for ambiguous ID cells, which both have high predictive uncertainty. The uncertainties are calculated using: $$
        \underbrace{\mathbb{H}\left[y |\mathbf{x}, \mathcal{D}\right]}_{\text{predictive}} = \underbrace{\mathbb{I}\left[y, \boldsymbol{\omega} |\mathbf{x}, \mathcal{D}\right]}_{\text {model}} + \underbrace{\mathbb{E}_{p\left(\omega |\mathcal{D}\right)}[\mathbb{H}[y |\mathbf{x}, \boldsymbol{\omega}]]}_{\text {data}},
$$
where $\mathbb{H}$ is the entropy and $\mathbb{I}$ the mutual information, $(\mathbf{x},y)$ the embedding and cell type of the cell, $\mathcal{D}$ the training data and $\mathbb{E}_{p\left(\omega |\mathcal{D}\right)}$ the expectation over the posterior \cite{smithUnderstandingMeasuresUncertainty2018}. The uncertainties reported throughout this work have been scaled to $[0,1]$.

In the context of label transfer from reference datasets, data uncertainty would apply to transitioning cell states which cannot be associated clearly with a single cell type. In contrast, model uncertainty highlights how familiar the model is with cells similar to the provided query cell, that is, in an ideal scenario, \emph{the model knows when it does not know}.

DKL and MIMO belong to the class of single forward pass uncertainty estimation models which we chose for their computational efficiency and performance. Other model classes that perform uncertainty estimation include: Bayesian Neural Networks \cite{blundell_BNN}, Bayesian Dropout \cite{gal_dropout}, or Ensemble models \cite{Lakshminarayanan}, but they are not further considered here.

We performed hyperparameter optimisation for all model classes across $\sim$100 distinct parameter combinations. Furthermore, all reported metrics are computed from model fits across three test sets evaluated on three different train-validation splits, resulting in 9 unique train-validation-test splits.

\subsection{Evaluation setup}
\label{sec:eval_setup}
As we are interested in both the evaluation of cell type transfer and the ability to provide meaningful uncertainty scores, the evaluation is divided in two parts. 

In section\,\ref{sec:part1}, we evaluate all models in terms of their balanced accuracy, F1-score and expected calibration error (ECE) \cite{guo2017calibration}. The leave-out set is a dataset integrated in the HLCA but not used during classifier training. Next to the predictive performance, we analyse the calibration of the classifiers. Calibration considers the relation between the frequency of correct predictions and the binned prediction probability and, therefore, indicates how reliable a classifier is. A well calibrated classifier will be correct 7 out of 10 times when providing a confidence score of $0.7$. For calibration evaluation we report both ECE and calibration curves.

In section\,\ref{sec:part2}, we test the models' ability to quantify uncertainty for several out-of-distribution (OOD) scenarios and their ability to discriminate between in-distribution training data (ID) and OOD data. To evaluate the models' uncertainty quality, we exclude three cell types during training and assess their performance through the predictive and model uncertainty. We evaluate OOD data discrimination using the Wasserstein distance \cite{Krishnan2020-dg} as well as the area under the precision-recall curve (AUPR).

\section{Results}
\label{results}
\begin{figure*}[ht]
    \vskip 0.02in
    \begin{center}
    \centerline{\includegraphics[width=0.8\textwidth]{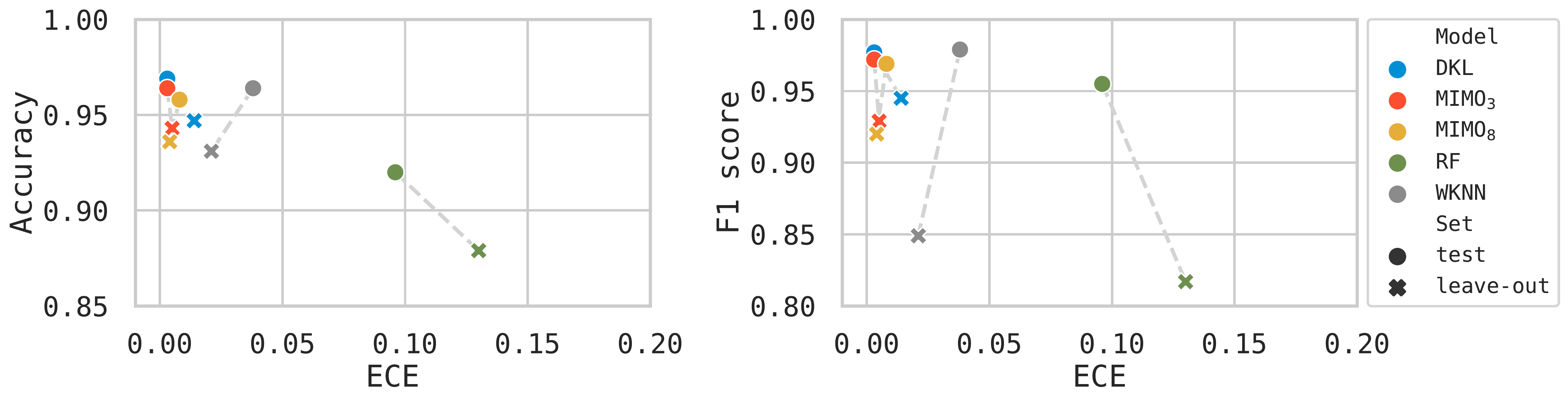}}
    \vskip -0.1in
    \caption{Accuracy ($\uparrow$) against ECE ($\downarrow$) \textit{(left)} and F1 score ($\uparrow$) against ECE ($\downarrow$) \textit{(right)} across all model classes. Each point represents the mean of three runs of nested cross-validation. The standard deviation is not reported as it is $<.002$ and does not affect the interpretation of the results.}
    \label{fig:acc_f1}
    \end{center}
    \vskip -0.3in
\end{figure*}

\begin{figure*}[ht]
    \vskip 0.02in
    \begin{center}
    \centerline{\includegraphics[width=0.8\textwidth]{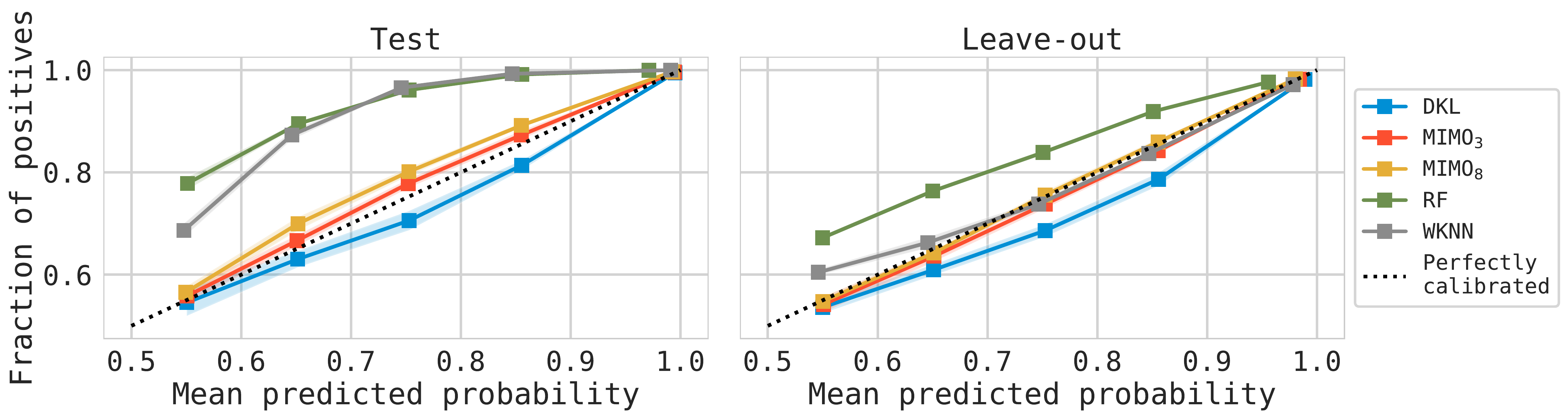}}
    \vskip -0.1in
    \caption{Fraction of positive predictions against the binned mean predicted probability across five bins for all five model classes. Each point represents the mean of three runs of nested cross-validation. The shaded area is the standard deviation.}
    \label{fig:calibration_curve}
    \end{center}
    \vskip -0.3in
\end{figure*}

\subsection{Label transfer and calibration}
\label{sec:part1}

We first evaluate the models on their ability to classify cell types on the integrated HLCA dataset using the most fine-grained cell type annotation (58 cell types in total). In a typical use case a new, leave-out dataset introduces a data-shift due to differences in experimental design, sample handling, and cell type composition. We use a random train-test split in this analysis to serve as upper bounds on the performance.

We find that all models have comparable accuracy and F1-scores (Figure\,\ref{fig:acc_f1}) on the test set. In contrast, the baseline models perform worse on the leave-out dataset, while DKL and MIMO are able to maintain their good performance. These results show that DKL and MIMO are superior in cell type classification to RF and WKNN.

The same is also reflected in the models' calibration, where the ECE for MIMO and DKL is consistently lower (better) compared to the baseline models. The calibration curves in Figure\,\ref{fig:calibration_curve} provide a more detailed view on calibration by showing the models' calibration at different confidence levels. It can be seen that WKNN and RF are further away from perfect calibration in the test set than DKL and MIMO. Interestingly, the decline in performance on the leave-out dataset leads to an improved calibration of WKNN, making it comparable to models that quantify uncertainty for the leave-out scenario. Nevertheless, this calibration plot, together with the ECE, should be seen in light of model accuracy, since a model should be robust not only in terms of calibration, but also accuracy under data shifts. We report exact scores in Tables\,\ref{tab:table1},\ref{tab:table2} in the appendix\,\ref{sec:appendix}.

Overall, this analysis indicates that models that quantify uncertainty, such as DKL and MIMO, are better suited for the cell type transfer task. They achieve comparable performance in terms of accuracy for both test and leave-out set, while also showing superior calibration compared to the WKNN and RF models.

\subsection{Uncertainty quantification on unseen cell types}
\label{sec:part2}

\begin{figure*}[ht]
    \vskip 0.05in
    \begin{center}
    \centerline{\includegraphics[width=0.8\textwidth]{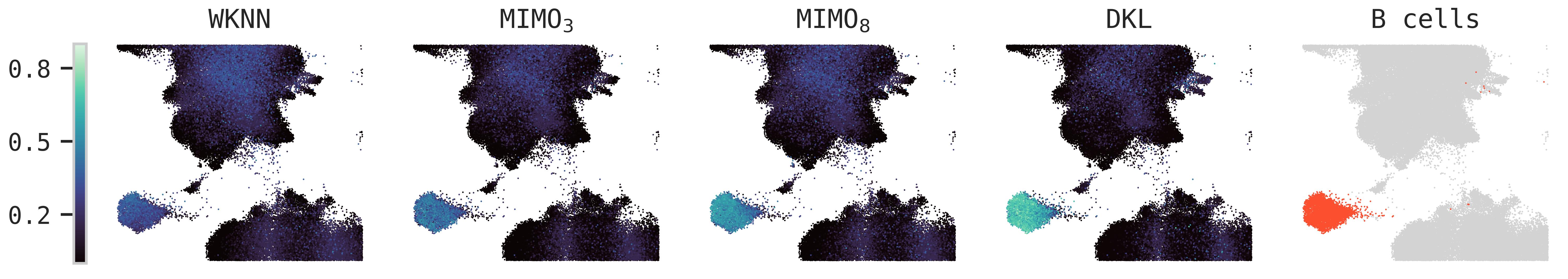}}
    \vskip -0.1in
    \caption{Predictive uncertainty on a subset of the HLCA UMAP with B cells left out during classifier training. It can be observed that the MIMO and DKL models predict higher uncertainty for the B cells. RF results are not reported due to poor classification accuracy.}
    \label{fig:umap_predictive_uncertainty}
    \end{center}
    \vskip -0.3in
\end{figure*}

\begin{table*}[ht]
\caption{Area under the precision-recall curve (AUPR) for classification between OOD (unseen cell type) and ID (validation set sampled from reference atlas), using different models and uncertainty types. DKL and MIMO outperform WKNN which is the baseline used in the HLCA. Softmax confidence uses the probability of the predicted class as a measure of uncertainty.}
\label{tab:aupr-ood}
\begin{center}
		\begin{small}
			\begin{sc}
				\begin{tabular}{llS[table-column-width = 1.9cm,table-format=2.0(0)]S[table-column-width = 1.9cm,table-format=2.0(0)]S[table-column-width = 1.9cm,table-format=2.0(0)]}
					\toprule
					\multirow[c]{2}{*}{Model}    & \multirow[c]{2}{*}{Uncertainty Type} & \multicolumn{3}{c}{AUPR (\%, pos. class: lo cell type)}                                                        \\
					{}                           & {}                                   & {b cells}                                               & {ionocytes}                   & {mast cells}         \\
					\midrule
					\multirow[c]{2}{*}{dkl}      & predictive                           & \bfseries\num{93+-3}                                    & \bfseries\num{94+-2}          & \num{95+-1}          \\
					                             & model                                & {}\num{82+-5}                                           & {\hspace{0.45em}}\num{70+-20} & \num{79+-1}          \\
					\midrule
					\multirow[c]{2}{*}{mimo$_8$} & predictive                           & \num{91+-1}                                             & {\hspace{0.45em}}\num{70+-10} & \num{94+-2}          \\
					                             & model                                & \num{93+-1}                                             & \num{82+-6}                   & \bfseries\num{96+-1} \\
					\midrule
					\multirow[c]{2}{*}{wknn}     & predictive                           & \num{61+-2}                                             & \num{24+-1}                   & \num{83+-1}          \\
					                             & softmax confidence                   & \num{51+-1}                                             & \num{12+-1}                   & \num{71+-2}          \\
					\bottomrule
				\end{tabular}
			\end{sc}
		\end{small}
	\end{center}
\vskip -0.1in
\end{table*}

To evaluate the models' ability to detect unseen cell types, we left out three cell types---B cells, Mast cells, and Ionocytes--- from the training data and compute the predictive uncertainty for all classes seen during training, as well as the left out class. A classifier trained on $K$ classes should result in high entropy of the output for the unseen, $K+1$ class.

Figure\,\ref{fig:umap_predictive_uncertainty} reports predictive uncertainty for B cells for all model types (see Figure\,\ref{supfig:predictive_umap} in appendix\,\ref{sec:appendix} for additional cell types). DKL clearly identifies B cells as a new, unseen cell type which is shown through the high uncertainty score for the respective observations. In contrast, predictive uncertainty for WKNN is diffused and does not highlight B cells. Consequently, WKNN's uncertainty estimates can be identified as less reliable which may cause incorrect cell type transfer. We quantitatively evaluate the densities of the predictive uncertainty for seen and unseen cell types via the Wasserstein distance, cf. \cite{Krishnan2020-dg}, see Figure\,\ref{supfig:wasserstein_distance} in appendix\,\ref{sec:appendix}, from which the same conclusions can be drawn.

Finally, we formulate the task of detecting a new cell type as a binary classification problem and quantify the performance through the AUPR. The results are shown in Table\,\ref{tab:aupr-ood}. Each entry is the AUPR of a model when the given uncertainty type is used for classification.
MIMO$_8$ performs best using model uncertainty for all left out cell types. DKL on the other hand has higher AUPR using predictive uncertainty. Since model uncertainty should be better suited for OOD detection, this suggests that MIMO$_8$ exhibits better disentanglement of model and data uncertainty. DKL's issue to highlight OOD cells using model uncertainty can also be observed in Figure\,\ref{supfig:model_umap} in appendix\,\ref{sec:appendix}. Nonetheless, DKL has higher AUPR than MIMO$_8$ for B cells and Ionocytes using predictive uncertainty. 
When using predictive uncertainty for DKL and model uncertainty for MIMO, both models outperform the WKNN baseline across cell types.

These results indicate that models that quantify uncertainty are superior in detecting unseen cell types, since they produce higher uncertainty on the left out classes. This is a desirable property of the models as it can be used by analysts to discover new cell states which are not present in the integrated scRNA-seq reference atlas.

This analysis requires further investigation and should be carried out on a real OOD scenario, i.e., where a new dataset is projected onto the reference and also excluded form the classifier's optimisation. Nonetheless, one has to appreciate that neither of the baseline models, RF and WKNN, provides such a measure on model uncertainty, and, therefore, an evaluation of models' prediction confidence is not possible. Put differently, it is not possible to state \emph{the model doesn't know when it doesn't know}. In contrast, models like DKL and MIMO, that provide measures on both data and model uncertainty, should be preferred for building and extending integrated scRNA-seq reference atlases.

\section{Conclusion}
\label{conclusion}
In this work, we presented uncertainty quantification on cell type transfer for single-cell reference atlases. We evaluated the considered models on a variety of metrics, including performance, calibration, and uncertainty quantification. We demonstrate that the baseline methods are not well calibrated and lack desirable robustness. In contrast, models that quantify uncertainty provide better confidence and more sensible uncertainty scores, which provide actionable results for downstream interpretations. In fact, such uncertainty measures can be used by analysts to uncover new cell types and cell states not present in the reference integrated atlas, as well as retain a notion on the ``confidence" of the prediction, so that more careful conclusions can be drawn. In the future, we seek to extend this benchmark by including real-world query-to-reference mapping as well provide a more extensive evaluation on model and data uncertainty under dataset shifts.\clearpage

\bibliography{references}
\bibliographystyle{icml2022}

\newpage
\appendix
\onecolumn
\section{Appendix}
\label{sec:appendix}
\setcounter{figure}{0}
\setcounter{table}{0}
\makeatletter
\renewcommand{\thefigure}{A\arabic{figure}}
\renewcommand{\theHfigure}{A\arabic{figure}}
\renewcommand{\thetable}{A\arabic{table}}
\renewcommand{\theHtable}{A\arabic{table}}

\begin{table}[!h]
\caption{Test set classification accuracy, F1 score and ECE}
\label{tab:table1}
\vskip 0.15in
\begin{center}
\begin{small}
\begin{sc}
\begin{tabular}{lrrr}
\toprule
& Accuracy ($\uparrow$) & F1 score ($\uparrow$) & ECE ($\downarrow$) \\
\midrule
DKL & 0.969 & 0.977 & 0.003 \\
MIMO$_3$ & 0.964 & 0.972 & 0.003 \\
MIMO$_8$ & 0.958 & 0.969 & 0.008 \\
RF & 0.920 & 0.955 & 0.096 \\
WKNN & 0.964 & 0.979 & 0.038 \\
\end{tabular}
\end{sc}
\end{small}
\end{center}
\vskip -0.1in
\end{table}

\begin{table}[!h]
\caption{Leave-out set classification accuracy, F1 score and ECE}
\label{tab:table2}
\vskip 0.15in
\begin{center}
\begin{small}
\begin{sc}
\begin{tabular}{lrrr}
\toprule
& Accuracy ($\uparrow$) & F1 score ($\uparrow$) & ECE ($\downarrow$) \\
\midrule
DKL & 0.947 & 0.945 & 0.014 \\
MIMO$_3$ & 0.943 & 0.929 & 0.005 \\
MIMO$_8$ & 0.936 & 0.920 & 0.004 \\
RF & 0.879 & 0.817 & 0.130 \\
WKNN & 0.931 & 0.849 & 0.021 \\
\end{tabular}
\end{sc}
\end{small}
\end{center}
\vskip -0.1in
\end{table}

\begin{figure*}[!ht]
\vskip 0.05in
\begin{center}
\centerline{\includegraphics[width=0.9\textwidth]{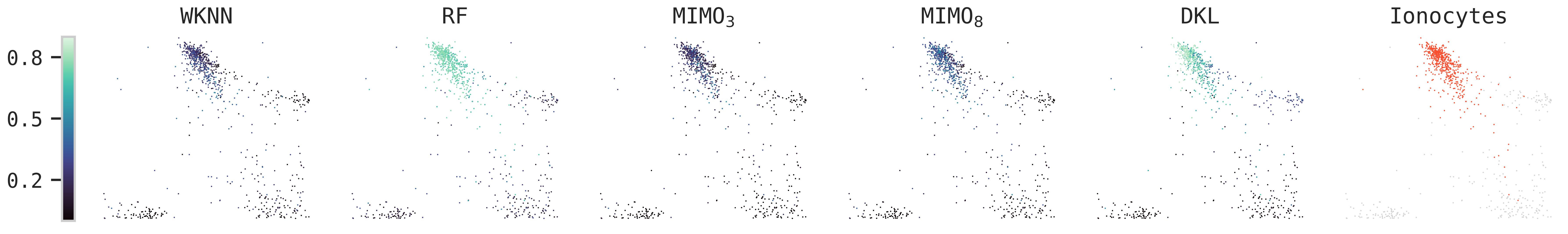}}
\centerline{\includegraphics[width=0.9\textwidth]{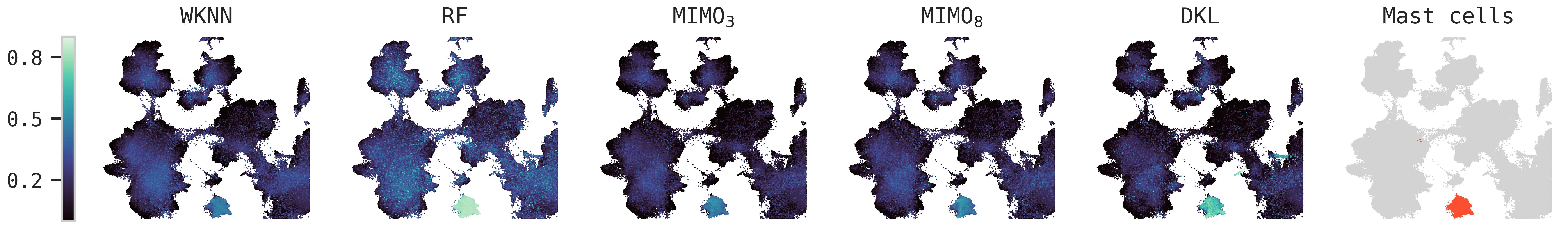}}

\caption{Predictive uncertainty for all models on left out cell types (rightmost panels).}
\label{supfig:predictive_umap}
\end{center}
\vskip -0.3in
\end{figure*}

\begin{figure*}[ht]
    \vskip 0.05in
    \begin{center}
    \centerline{\includegraphics[width=0.9\textwidth]{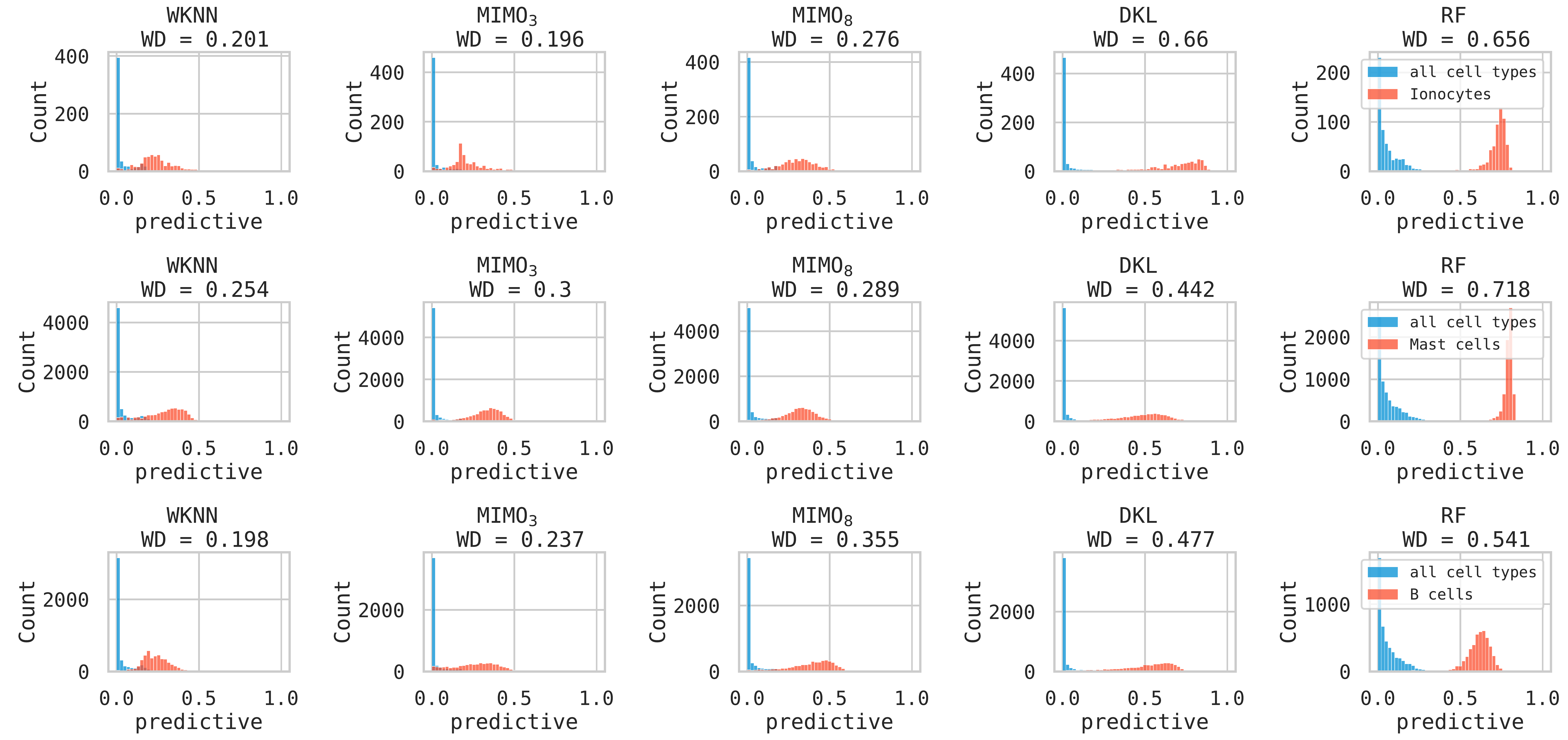}}
    \vskip -0.1in
    \caption{Distribution of predictive uncertainty for the cell types the model was trained on (blue) and the left out cell type (orange). The  computed Wasserstein distance, WD, ($\uparrow$) between both distributions is shown in the title. This quantifies the models' ability to distinguish the OOD cell type from the ID training data. All models outperform the WKNN baseline.  While RF shows the largest WD, using this model cannot be recommended due to poor performance in classification accuracy.}
    \label{supfig:wasserstein_distance}
    \end{center}
    \vskip -0.3in
\end{figure*}

\begin{figure*}[!ht]
\vskip 0.05in
\begin{center}
\centerline{\includegraphics[width=0.9\textwidth]{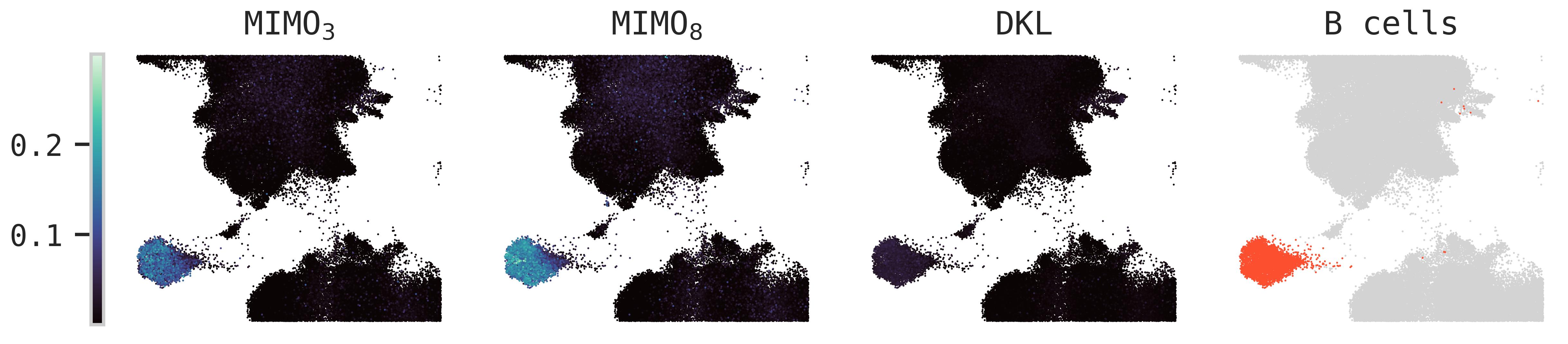}}
\centerline{\includegraphics[width=0.9\textwidth]{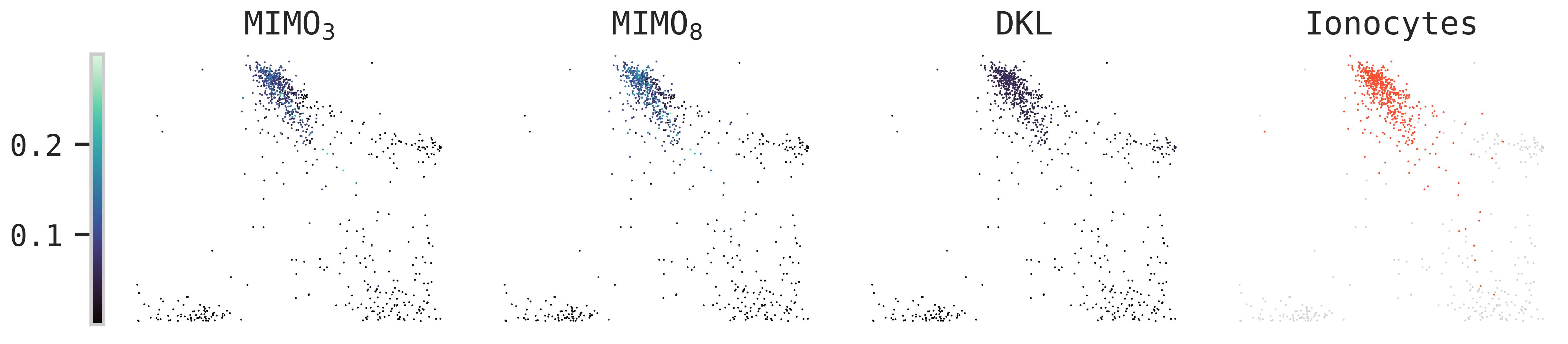}}
\centerline{\includegraphics[width=0.9\textwidth]{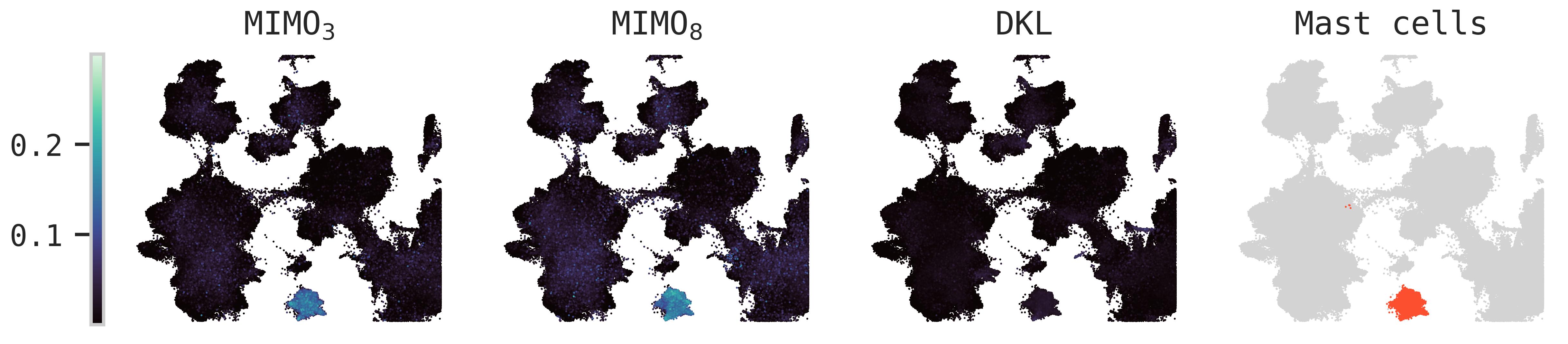}}

\caption{Model uncertainty for MIMO and DKL on left out cell types (rightmost panels).}
\label{supfig:model_umap}
\end{center}
\vskip -0.3in
\end{figure*}

\begin{figure*}[!ht]
\vskip 0.05in
\begin{center}
\centerline{\includegraphics[width=0.65\textwidth]{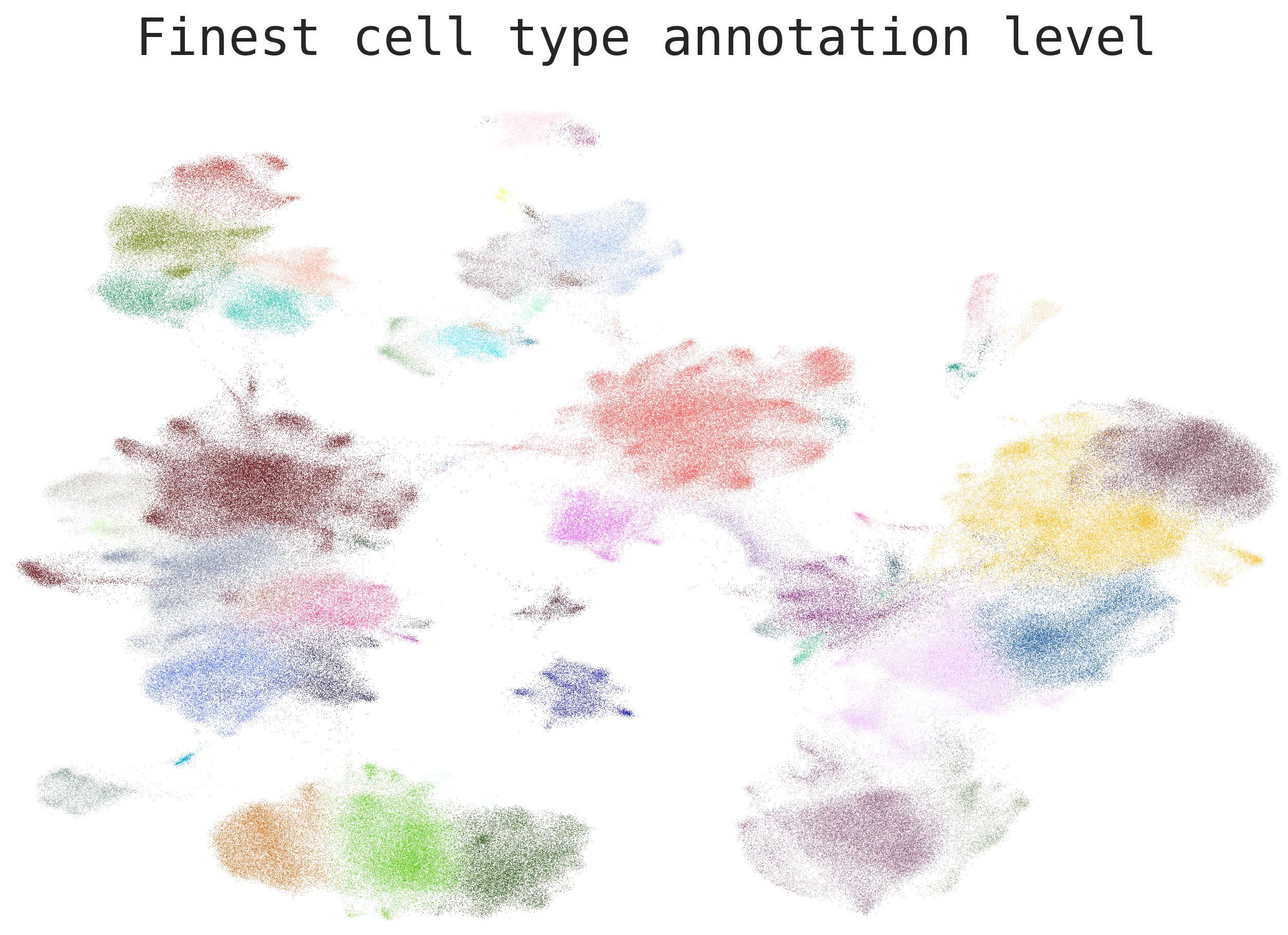}}
\vskip 0.05in
\centerline{\includegraphics[width=0.55\textwidth]{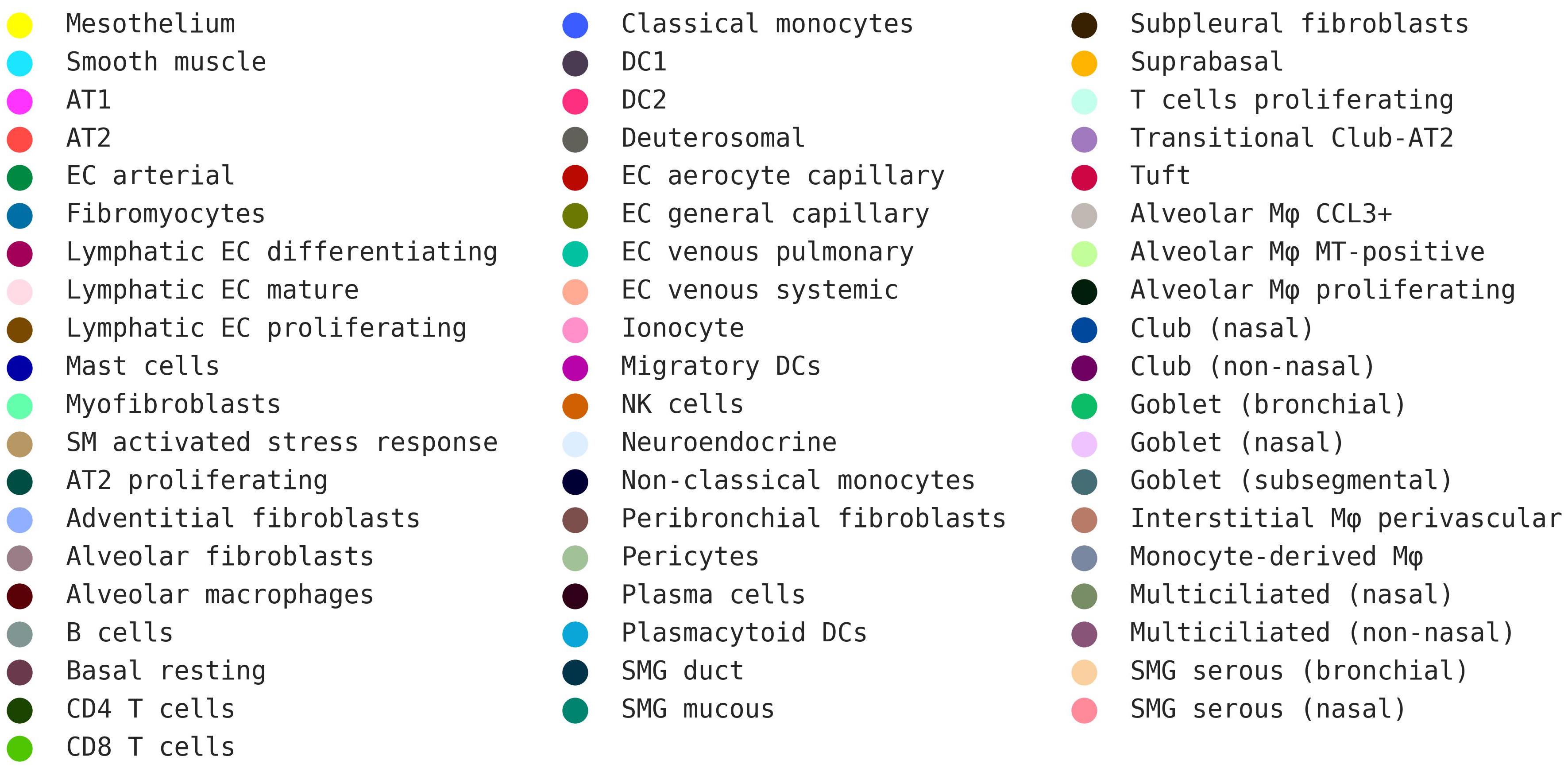}}
\vskip 0.25in
\centerline{\includegraphics[width=0.7\textwidth]{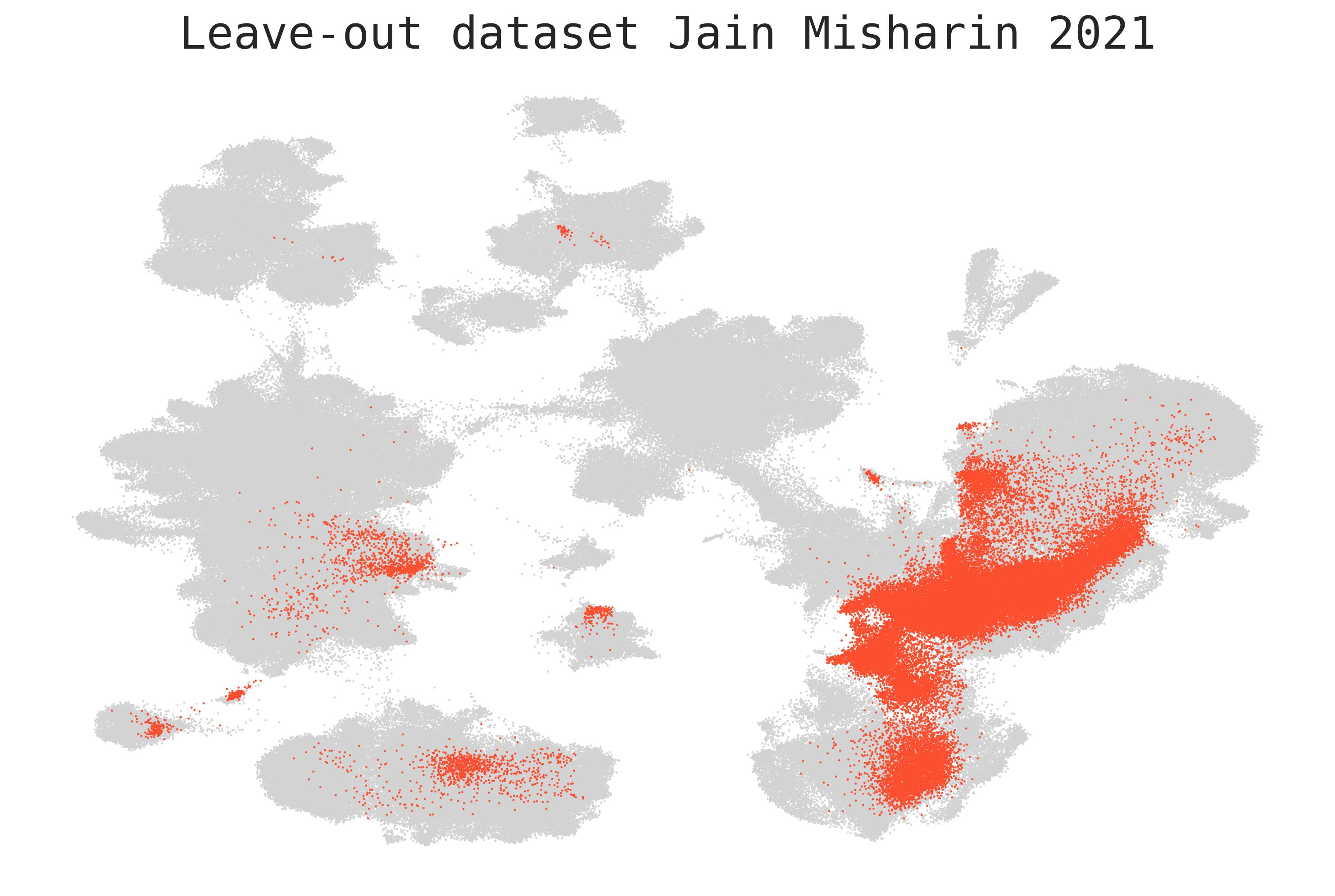}}

\caption{Human Lung Cell Atlas UMAPs colored by cell types (top) and leave-out dataset (bottom).}
\label{supfig:hlca_umap}
\end{center}
\vskip -0.3in
\end{figure*}

\end{document}